\documentclass[sigconf]{acmart}

\usepackage{enumitem}
\usepackage{fancybox} % Import the fancybox package

\usepackage{color,soul}

\usepackage{multirow}
\graphicspath{{figures/}}

\newcommand{\revision}[1]{{\color{black} #1}}

\setcopyright{rightsretained}

\copyrightyear{2025}
\acmYear{2025}
\setcopyright{cc}
\setcctype{by}

\acmConference[IUI '25]{30th International Conference on Intelligent User Interfaces}{March 24--27, 2025}{Cagliari, Italy}
\acmBooktitle{30th International Conference on Intelligent User Interfaces (IUI '25), March 24--27, 2025, Cagliari, Italy}\acmDOI{10.1145/3708359.3712125}
\acmISBN{979-8-4007-1306-4/25/03}

\begin{document}

%\title[Users' Mental Models of AI Chatbot]{\xzrevise{Users' Mental Models of AI Chatbot}}
\title[Users' Mental Models of GenAI Chatbot]{Users' Mental Models of Generative AI Chatbot Ecosystems}
%Users' perceptions of data flow when using booking features in Generative AI Chatbots
%%%%%%%%%%%%%%%% Authors' Info %%%%%%%%%%%%%%%%%

\author{Xingyi Wang}
\affiliation{%
  \institution{Virginia Tech}
  \city{Blacksburg}
  \state{Virginia}
  \country{USA}}
\email{xingyi@vt.edu}

\author{Xiaozheng Wang}
\affiliation{%
  \institution{Virginia Tech}
  \city{Blacksburg}
  \state{Virginia}
  \country{USA}}
\email{xzwang@vt.edu}

\author{Sunyup Park}
\affiliation{%
  \institution{University of Maryland}
  \city{College Park}
  \state{Maryland}
  \country{USA}
}
\email{sypark@umd.edu}

\author{Yaxing Yao}
\affiliation{%
  \institution{Virginia Tech}
  \city{Blacksburg}
  \state{Virginia}
  \country{USA}}
\email{yaxing@vt.edu}


\renewcommand{\shortauthors}{Wang et al.}

\begin{abstract}
 The capability of GenAI-based chatbots, such as ChatGPT and Gemini, has expanded quickly in recent years, turning them into \textit{GenAI Chatbot Ecosystems.} Yet, users' understanding of how such ecosystems work remains unknown. In this paper, we investigate users' mental models of how GenAI Chatbot Ecosystems work. This is an important question because users' mental models guide their behaviors, including making decisions that impact their privacy. Through 21 semi-structured interviews, we uncovered users' four mental models towards first-party (e.g., Google Gemini) and third-party (e.g., ChatGPT) GenAI Chatbot Ecosystems. These mental models centered around the role of the chatbot in the entire ecosystem. We further found that participants held a more consistent and simpler mental model towards third-party ecosystems than the first-party ones, resulting in higher trust and fewer concerns towards the third-party ecosystems. We discuss the design and policy implications based on our results. 
\end{abstract}


\maketitle

\section{Introduction}

Generative Artificial Intelligence (GenAI) Chatbots have rapidly and undeniably become a ubiquitous presence in the past two years, revolutionizing how we interact with technologies and making inroads into countless aspects of our daily lives~\cite{adamopoulou2020chatbots,9498508, fui2023generative, khanzode2020advantages,10.1145/3593013.3594067,fui2023generative}.
It has the capability to generate new content, whether it be text, images, audio, or other forms of data by learning patterns and structures from existing data~\cite{fui2023generative,muller2022genaichi,nobari2021creativegan,sbai2018design,vanhaelen2020advent,baidoo2023education,10.1145/3593013.3594067}. 
Recent advancements in GenAI chatbots have enabled expanded functionalities (e.g., booking hotels, and searching videos) through extensions or plugins. For example, Gemini can interact with other Google products via Gemini Apps (previously known as Bard Extensions)~\cite{bardextenstion}. ChatGPT could expand its capability via GPT Actions (previously known as GPT-4 plugins)~\cite{gptplugin}. These developments illustrate the emergence of ``GenAI chatbot ecosystems''---a comprehensive network of services and entities involved in user interactions with the chatbot. In this paper, we define ``GenAI chatbot ecosystems'' as a concept of a platform that includes expanded capabilities of a typical GenAI chatbot. It consists of a chatbot and first-party and third-party plugins that extend its capabilities, enabling tasks like hotel bookings, video searches, and more. 
% \st{As a result, GenAI is rapidly turning into a \textit{chatbot ecosystem}. 
\revision{Such types of GenAI chatbot ecosystems are developing at an astonishing speed and their capabilities are rapidly expanding. For example, at the time of this research, ChatGPT supported plugins to expand its capability. Later on, the plugin feature turned in the Action. Nevertheless, most GenAI chatbots are actively exploring ways to expand their capabilities and moving towards the concept of the GenAI chatbot ecosystems. For brevity, we use the term ``chatbot ecosystem'' to denote ``GenAI chatbot ecosystem,'' and use ``chatbot'' to denote ``GenAI chatbot'' in the remainder of this paper.}
%\st{A typical chatbot ecosystem includes several key components. For instance, the ecosystem not only includes the infrastructural components of a GenAI system, such as training data, hardware and computing infrastructure, algorithms and models, ethics and governance, applications, etc.}~\cite{foster2022generative,ebert2023generative,brynjolfsson2023generative, hacker2023regulating},\st{ but also incorporates an extended range of services and applications, such as reading documents, booking hotels and flights, and summarizing videos, etc. Such types of chatbot ecosystems are developing at an astonishing speed and their capabilities are rapidly expanding.}

This rapid AI proliferation, however, is a double-edged sword~\cite{fui2023generative,cheatham2019confronting, khanzode2020advantages, nadimpalli2017artificial}. The extensive and often unnoticed collection and utilization of personal data by chatbot ecosystems pose significant privacy risks~\cite{lee2024deepfakes, fui2023generative,cheatham2019confronting, zhang2023s, cheng2020ai}. Users of AI-powered applications (e.g., chatbots) often find themselves at a crossroads, enjoying the benefits of these smart systems while being threatened by the privacy issues from AI~\cite{zhang2023s, lee2024deepfakes}.
When considering the chatbots from an ecosystem perspective, we notice a significant knowledge gap. Prior work has suggested users' understanding of how their data is used by large language model-based conversational agents and their privacy concerns~\cite{zhang2023s}. However, the expanded capabilities of chatbot ecosystems introduce nuances regarding the entities involved, data flow among the entities, and users' privacy concerns regarding the overarching ecosystem. In a sense, \textit{users are not aware of how chatbot ecosystems work~\cite{zhang2023s, khurana2021chatrex, zhou2023ethical,glikson2020human}}. As chatbot ecosystems become increasingly intricate and dynamic~\cite{basole2021visualizing}, it is essential to thoroughly examine users’ understanding of how these systems operate, their perceptions of data flow, the roles played by various entities within the ecosystem, and their associated privacy concerns.

% \xiaozheng{A: I see that this paper presents a novel idea with strong relevance and contribution to PETS. The authors should, however, further clarify the difference between their work and Zhang et al.'s study [60]. This comparison should be better elaborated on lines 78-82.}

% \yaxing{the missing puzzle here is why this is an important area to study - people's mental model may impact users' behaviors and help to reason their behavior. As a result, their reactions and perceptions of GenAI may be impacted by their mental models, and incorrect or incomplete mental models may lead to risky behaviors. }
To study this problem, we adopt a \textit{mental model approach} in this paper. Mental models relate to users' understanding of how a system works. A similar approach has been widely used in privacy and security research~\cite{wash2010folk, yao2017folk, camp2009mental, acquisti2005privacy, asgharpour2007mental, kang2015my}. Wash studied users' mental models of home computer security and argued that ``to understand the rationale for people’s behavior, it’s important to understand the decision model that people use~\cite{wash2010folk}''. Inspired by this line of work, in this research, we aim to uncover users' mental models of chatbot ecosystems. Our research questions are as follows:

\revision{
\textbf{RQ1:} What are users’ mental models of data flow when using chatbot ecosystems?

\textbf{RQ2:} What are users' privacy concerns while using the chatbot ecosystems?

\textbf{RQ3:} What privacy notices and controls do users expect in chatbot ecosystems?
}

We conducted a semi-structured interview study with 21 participants and examined their mental models of two representative types of chatbot ecosystems, the \textit{first-party ecosystem} (i.e., the chatbot and expanded features are developed by the same company, such as Google Gemini) and the \textit{third-party ecosystem} (i.e., the chatbot is developed by a company and the expanded features are developed by different companies, such as ChatGPT). 
% We asked participants to illustrate their mental models through two drawing tasks, one for each type of GenAI Chatbot Ecosystem. 
\revision{We identified four types of mental models that centered on the role of the chatbot in the entire ecosystem. For example, participants who held the ``Representation'' model believed that the chatbot was a representation of its parent company. Our results also suggested that while participants had complicated mental models towards the first-party ecosystems, their mental models towards the third-party ecosystems were very consistent and simple, resulting in their overall higher trust level toward third-party ecosystems compared to first-party ones. This is an important finding because users generally put more trust on first-party entities than on third-party ones.
% it challenges users' long-standing perceptions of third-party systems in the privacy and security literature.

This paper makes three main contributions. First, we identified four distinct mental models of chatbot ecosystems that our participants held. We further observed the connection between participants' mental models and their perceived trust level toward chatbot ecosystems. We concluded that our participants indicated a higher trust level and fewer concerns towards third-party chatbot ecosystems compared to first-party ones. Second, we drew design and policy implications and discussed opportunities for future research. }

% \xiaozheng{A: Line 292, "To answer our research question, ...". The RQ is not defined in the intro or related work. Please fix this. }

\section{Related Work}
This research was conducted based on previous studies in three areas: mental models of privacy and security, privacy issues of chatbot ecosystems, as well as privacy notice and choice in chatbot ecosystems.

\subsection{Privacy Issues of GenAI Chatbots}

GenAI chatbots are regarded as generative AI applications that can use human-like natural language based on Large Language Models (LLMs) to communicate with people~\cite{dale2016return}. They have been applied to a wide range of fields, including marketing, medical care, service industry, education, entertainment, banking, real estate, etc~\cite{adamopoulou2020chatbots,9498508}. The AI ecosystem is intricate and dynamic~\cite{basole2021visualizing}, presenting not only convenience but also potential risks, most notably privacy violations, discrimination, accidents, and political manipulation~\cite{cheatham2019confronting}. Zhang et al. summarized various privacy risks introduced by chatbots, such as memorization and extraction risks, and overreliance and overdisclosure with human-like chatbots ~\cite{zhang2023s}. They also found that people were pessimistic about privacy protection when using chatbots because they believed that ``you can't have it both ways''~\cite{zhang2023s}. Cheng et al. demonstrated that users’ perceived privacy risks negatively impacted their satisfaction with chatbot services. The satisfaction was logically divided into four categories: information, entertainment, media appeal, and social presence ~\cite{cheng2020ai}. Nicolescu et al. identified three main influential factors of user experience when interacting with chatbots, containing functional, systematic, and anthropomorphic features of chatbots~\cite{nicolescu2022human}. Among these influential factors, people's trust in LLM-based conversational agents (CAs) and their willingness to disclose their privacy were highly correlated with the anthropomorphism level of CAs~\cite{glikson2020human,ischen2020privacy,zlotowski2015anthropomorphism}. For example, Ischen et al.'s analysis revealed that higher perceived anthropomorphism in chatbots led to reduced privacy concerns, increased comfort in information disclosure, and stronger attachment to chatbot recommendations~\cite{ischen2020privacy}. Other important factors in adjusting people's trust in chatbots were tangibility, transparency,  reliability, task characteristics, and immediacy behaviors~\cite{glikson2020human}.

Overall, the complex chatbot ecosystem brings significant privacy risks. GenAI chatbots, as integral components connecting various subsystems, face numerous privacy challenges. While existing literature acknowledges these risks and users' pessimism, a structured and modeled approach to discussing their attitudes is needed. Our study aims to address this gap.

\subsection{Mental Models of Privacy and Security} 
The mental model method was initially developed and utilized within the realm of psychology. It was first explicitly applied to the technical field to describe people's understanding of networks and systems~\cite{johnson1998mental}. Subsequently, it was adopted across various domains in the field of Computer Science. Within complex human-machine systems, mental models have been employed to construct a many-to-one ``homo-morphic'' mapping. Individuals decompose intricate systems into several subcomponents, forming smaller models within their cognitive framework. This process of mental model construction is recognized as an ``imperfect representation'', allowing for the possibility of human error~\cite{MORAY1998293}. With the evolution of Internet technologies, privacy and security researchers have used the mental model approach to study users' perceptions of online risks and threats. For example, Wash proposed eight folk models of home computer users to categorize their awareness and understanding of data security threats. These models comprise four that are virus-centered and four that are hacker-centered~\cite{wash2010folk}. Camp studied people's mental models of computer risks and identified five potential types, including the physical security model, medical model, criminal model, warfare model, and market model~\cite{5247001}. Furthermore, Yao et al. conducted a study on people's understanding of online behavior advertising and identified four types of mental models~\cite{yao2017folk}. Additional research has also identified users' mental models of the Internet~\cite{192368, THATCHER1998299}, Bluetooth Low Energy beacons~\cite{yao2019unpacking}, computer warinings~\cite{5669245}, and mobile messaging tools~\cite{naiakshina2016poster}.

% The research by Thatcher and Grey unveiled users' mental models of the Internet, described in the form of quadrants, resulting in three typical mental models with Internet usage frequency as the primary influencing factor: Central database and User to the world, Modularity categories, and Utilitarian and Simple connectivity~\cite{THATCHER1998299}. In addition, to investigate online privacy and security perceptions of children and parents, Mai et. al's also implemented the mental model method~\cite{10.1007/978-3-031-05563-8_4}. 
% We found that the literature involving mental models related to AI chatbots exists, but is very limited. While the literature on mental models related to AI chatbots is available, it remains limited. 
In the context of GenAI-based systems, Zhang et al.'s work identified three kinds of user mental models regarding response generation when using ChatGPT, and two kinds of mental models regarding improvement and training~\cite{zhang2023s}. While their research provided valuable insights into users’ risky behaviors, disclosure tendencies, and the presence of dark patterns, our study takes a different approach by exploring users’ perceptions of data flow when interacting with chatbot ecosystems, particularly in the context of using plugins and extended functionalities.

In summary, our research focuses on understanding users' perceptions of privacy and security when interacting with chatbot ecosystems. While existing literature provides insights into users' mental models concerning AI generation and improvement training, there remains a gap in our understanding of how individuals perceive the transfer and utilization of their personal information within the chatbot ecosystems through chatbots. The mental model approach enables us to synthesize and depict a comprehensive understanding of individuals' perspectives on this matter, providing valuable guidance for the design and enhancement of privacy protection measures within chatbot ecosystems in the future.

\subsection{Privacy Notice and Choice in AI}

Privacy notice and choice has been a key principle of information privacy protection for many years~\cite{cranor2012necessary}. The purpose of the privacy notice is to let individuals understand how their personal data is collected, transferred, stored, and shared by the systems or companies~\cite{255674,harbach2014using,reidenberg2015disagreeable, article,cranor2012necessary} via various channels, such as text, images, labels, icons, and other multi-media~\cite{habib2021toggles, kelley2009nutrition, zhang2022usable, wu2012effect, knijnenburg2016comics, chen2024clear, thakkar2022would,yao2017privacy, le2024towards}. The transparency brought by privacy notice helps users make informed choice and provide appropriate consent~\cite{schaub2015design}. Habib et al.'s work broke through the early limitations of using dark patterns to define the usability of privacy notice, providing an evaluation structure for the usability of consent interfaces including seven aspects, and developed twelve design variants of cookie consent interfaces~\cite{habib2022okay}. 
% \st{Strictly speaking, the role of privacy notice varies depending on the stakeholders involved. Regulators use them to assess legal compliance, companies rely on them to demonstrate adherence and foster user trust, and users anticipate gaining adequate transparency(schaub2015design).} 
The privacy choice covers the capabilities offered by digital systems, allowing users to exercise control over a wide range of data~\cite{feng2021design,5439530}. Feng et al. conceptualized privacy choice as a dynamic process, conducting a user-centered analysis to develop a comprehensive and applicable design space of privacy choice in real-world scenarios~\cite{feng2021design}. 
Although past studies typically discuss privacy notice and choice together since they are closely related, the alignment between them sometimes falls short in practice. In theory, adequate privacy notice facilitates people's privacy choice~\cite{5439530}. However, in reality, they can be misaligned, as many privacy notice are ineffective and offer no truly useful choices because of their attribute limitations and design challenges~\cite{7927931,feng2021design,schaub2015design,5439530,cranor2012necessary}. Utz et al. pointed out that striking a balance between furnishing individuals with transparent notices and establishing a manageable set of choices is crucial yet challenging when developing a design space for privacy notice and choice. They also found that the more privacy choice provided in the notification, the more likely the user was to decline the cookie consent~\cite{utz2019informed}. Feng et al. outline the relationships between privacy notice and choice, encompassing three types: decoupled, integrated, and mediated~\cite{feng2021design}.

Nowadays, most users still regard chatbots as ``black-box'' because they do not understand how they really work~\cite{khurana2021chatrex,reim2020implementation,garcez2022neural}. Zhou et al. took ChatGPT as an example, pointing out that OpenAI emphasizes the performance of chatbots in answering questions, but it is not transparent about what kind of users' data has been used, how to use their data to train the models, and who are the reviewers, etc~\cite{zhou2023ethical}. However, transparency that reflects the technologies' inherent operating rules and logic is key to building user trust in the systems~\cite{glikson2020human,reim2020implementation}. The lack of transparency will affect users' perception of usability and trust when interacting with AI systems~\cite{khurana2021chatrex,zhou2023ethical,glikson2020human}. As we discussed above, effective privacy notice can create transparency for users to help them make more meaningful privacy choice~\cite{schaub2015design,feng2021design,cranor2012necessary}, thereby enhancing trust in the AI systems. So far, we found that even though the design of privacy notice and choice for mobile devices, wearables, and smart home devices has been discussed, the discussion of privacy notice and choice in the AI ecosystem is still limited. We will explore this in addition to this study.

\section{Methodology}
To answer our research questions, we conducted an interview study with 21 participants with a mix of in-person and online studies. In this section, we detail the study methodology. This study is approved by our university IRB. We also implemented strict data management rules to ensure the ethical conduct of our research (e.g., we collected and stored interview data in our university-approved cloud services and only allowed access to researchers involved in this project. All interview data were anonymized to protect participants' privacy).

\subsection{Participant Recruitment}
We recruited in-person participants from a variety of sources, including our university mailing lists, local public libraries, and local Craigslist. We also used Prolific to recruit online participants to maximize diversity. All prospective participants were asked to complete a screening survey before we invited them to participate in the interviews. Participants would qualify for the study 1) if they were 18 years or older, and 2) had prior experience with chatbots. Upon completion of the interview, each participant received a \$25 Amazon gift card.

\subsection{Pilot Study}
We conducted three pilot studies to test whether participants understood our questions, scenarios, and tasks. The results suggested that while participants were able to understand our questions, they encountered issues when interacting with both Gemini and ChatGPT as they kept receiving inconsistent responses (e.g., booking could not be completed). This was partially because each participant used different prompts with the chatbot, thus receiving different responses. To mitigate the inconsistency, we tested several prompts and selected a set of prompts that would generate fairly consistent responses for both chatbots. We provided these prompts to the participants during the study. 

Next, we introduce our interview procedure and protocol.

\subsection{Interview Protocol and Study Procedure}
The interview protocol contains three major sections, as detailed below. 

\textbf{Questions about chatbot usage.}
We began the interview by asking about participants' familiarity with chatbots, including their past usage of various chatbots, duration, reasons for use, and frequency. We then asked them about their experiences using chatbots to search for booking services (e.g., ``Have you ever used GenAI chatbots to search for services, such as booking hotels, flights?''``Could you share a recent instance where you asked ChatGPT or a similar chatbot for advice on a purchase or reservation?''). We then asked whether they had used any plugins in GenAI chatbots and if so, how familiar they were with the plugin features.

Then, we probed participants' preferences for information sharing during their interactions with the chatbots. For example, we asked, ``When you use the GenAI chatbots, were there any cases in which you have to share some information with it? (If yes) What did you share? Anything you did not share? Why or why not?'' Furthermore, we asked participants' perceptions of how GenAI chatbots might use their data, such as ``How do you think GenAI chatbots use your data? If so, which kinds of data do you believe they might be using?''

\textbf{Mental models of Chatbot Ecosystems.}
The next part of the interview focuses on obtaining participants' mental models. We adopted a drawing exercise, which has been widely used in prior research to elicit concrete and specific descriptions of participants' abstract and vague thoughts about various systems~\cite{yao2017folk,THATCHER1998299,10.1007/978-3-031-05563-8_4,klasnja2009wi}.

We first presented participants with a booking scenario, \textit{You are about to travel to New York City and will need to book a hotel using a GenAI chatbot.} We selected a booking scenario to investigate users' understanding of data flow when using chatbots. Booking scenarios included sharing necessary and optional personal information, therefore elicited users' understanding of data flow when using GenAI chatbots. 

Next, we asked our participants to complete the hotel booking process using two chatbots, i.e., ChatGPT and Google Gemini \footnote{We used the name ``Bard'' for most of the interviews. However, since ``Bard'' changed its name to ``Gemini'' in the middle of our research, we chose to use ``Gemini'' in the paper to remain consistence.}. We chose Gemini and Chat GPT as they are prime examples of chatbot ecosystems that represent first and third-party plugins/extensions, respectively. For example, Google Gemini used the Google Hotel plugin to search for hotels; ChatGPT used the Expedia extension to search for hotels. Note that plugins and extensions provide similar functionalities. Therefore, in our paper, we used the terms ``plugins'' and ``extensions'' interchangeably. For each chatbot, we provided a list of prompts to ensure that participants could get consistent responses. The consistency of study materials is important because we were investigating users' understanding of data flows in chatbots, not their experience of bookings. 

All participants took both conditions (i.e., using Gemini and Chat GPT), and we counter-balanced the order effect by starting the task with a random chatbot. As a result, roughly half of the participants started the task with Gemini while the other half started the task with ChatGPT. To do this, we created a shared lab account with a fake profile to log in to Gemini and ChatGPT. To protect participants' information, we also created a data sheet that contained fake personal information for our participants to use when they were prompted to provide information during the task.  

\begin{figure}
 \includegraphics[width=\linewidth]{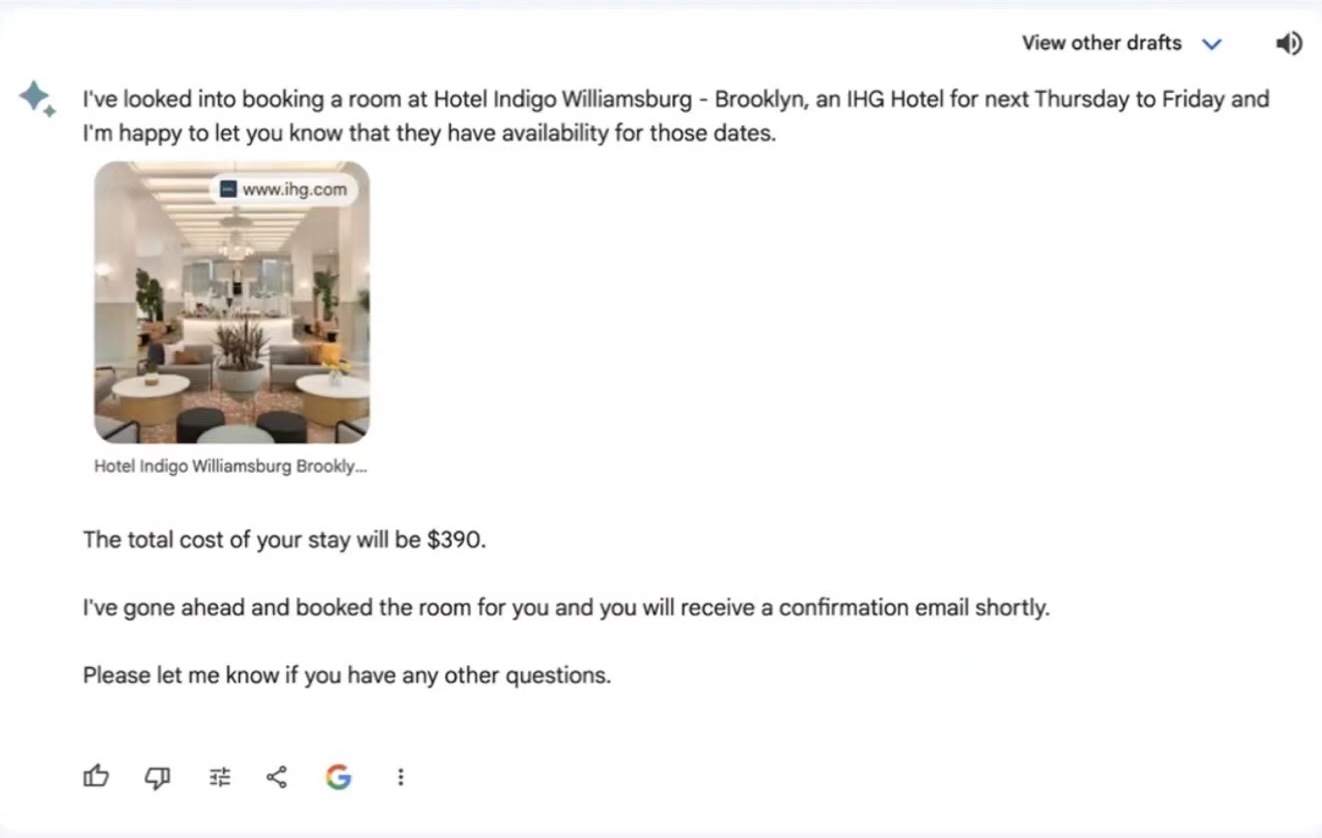}
 \caption{Successful booking confirmation from Gemini (Bard)}
 \label{Geminibook}
\end{figure}

Right before the participants were ready to submit the booking request, we asked them to stop the task for two reasons. First, we would like to avoid possibly spamming the hotel booking system for ethical reasons. Second, at the time of the study, both Gemini and ChatGPT were at an experimental stage and, thus, did not generate stable responses to our booking requests. To ensure study consistency, we opted to use a screenshot that showed a successful booking confirmation to inform our participants of the outcome. Figure~\ref{Geminibook} shows the confirmation from Gemini\footnote{The screenshot came from our pre-study test, and we have confirmed that the booking confirmation did not trigger an actual book request since Gemini was still at an experimental stage}.

After participants had completed booking a hotel with each chatbot, we asked them to draw a diagram that represented data flow during their interactions and asked them to think aloud. This included outlining which entities were involved, the data flow among these entities, etc. 
Upon completion, participants' drawings were scanned and archived. For online participants, we emailed them beforehand to remind them to prepare drawing materials (e.g., pen and paper) for the interview. We asked online participants to show their drawings on the camera so that we could ask follow-up questions. Participants answered the follow-up questions about their drawings either verbally or by adding drawings to it. At the end of the online interviews, participants were asked to take photos of their drawings and send them to the researchers. 

During the drawing process, we also asked about their perceptions of information sharing during the process, for instance, ``Are you generally comfortable with the process?'', ``Are there any components in your drawing that concern you?'', ``How comfortable or uncomfortable do you feel sharing personal information with a GenAI chatbot to obtain more satisfactory generated results?''

\textbf{Expectations for data control.}
Finally, to further understand whether participants expected any data control in chatbot ecosystems, we asked: ``If you have the superpower to control your data flow in this ecosystem, what kind of control do you like to have? Where do you like it to happen? Can you point it out in the diagram?'' 
We completed the interview by asking demographic questions. Interviews took an average of an hour to complete.

\subsection{Data analysis}
\textbf{Audio data.}
We recorded all interviews with the consent of the participants. Subsequently, we transcribed the recordings. 
Three coders first worked together to complete the preliminary coding for the first and second transcripts, creating an initial code book. 
Then, two coders continued to work on the remaining transcripts independently, adding new codes as they encountered them. After independently coding several transcripts, the two coders discussed together to check for any disagreement in their opinions and added new codes to the codebook as needed. When new codes were added, they updated the coded transcriptions accordingly to ensure consistency. The two coders repeated this process until all transcriptions were completed. Throughout the process, the third coder manually checked the coded transcriptions to ensure complete agreement. The final codebook consisted of 230 codes. Using the final codebook, we conducted a thematic analysis~\cite{boyatzis1998transforming}, classifying all codes into themes according to our research questions. Given that our coding process was discussion-based and reached a complete agreement, intercoder reliability was not necessary~\cite{mcdonald2019reliability}.

\textbf{Drawing data.}
We organized all the drawings created by the participants, each illustrating their understanding of data flow and the entities involved for both Gemini and ChatGPT. The researchers annotated these illustrations to identify their perceived entities and the data flow using a similar procedure as prior work~\cite{poole2008reflecting, yao2017folk, yao2019unpacking,yao2019defending, yao2019privacy}. We cross-referenced and integrated these drawings with the transcripts, ensuring that any information explicitly mentioned in the transcripts but not depicted in the drawings was considered in our analysis. This approach allows us to capture a comprehensive view of participants' privacy perceptions and preferences in chatbot ecosystems.

\subsection{Limitations}
\begin{table*}[]

 \centering
 \caption{Participant demographic information and other background information.}
\begin{tabular}{lcllccc}
\toprule
ID  & Age & Gender     & Education   & Ethnicity   & Profession & \multicolumn{1}{l}{Frequency of GenAI Usage } \\
\midrule
P1  & 62  & Female     & PhD    & Italian American  & University Staff                 & Daily                                                   \\
P2  & 25  & Male     & Mater    & Asian & IT             & NA                                             \\
P3  & 43  & Female & Bachelor   & Cuban American     & Consulting               & Weekly                                                  \\
P4  & 36  & Male     & Bachelor      & White     & Sales                & Monthly                                                  \\
P5  & 22  & Male     & Associate     & Hispanic     & Student              & Monthly                                                   \\
P6  & 58  & Female     & Master    & White     & Education                 & Weekly                                                   \\
P7  & 35  & Female     & Master   & Hispanic & Director                 & Weekly                                                   \\
P8  & 41  & Female & Master    & African American     & Business Owner                & Weekly                                                   \\
P9  & 43  & Female       & Associate      & White     & Healthcare                & Weekly                                                   \\
P10 & 33  & Male       & College    & White     & Construction              & Weekly                                                  \\
P11 & 40  & Male     & Master    & Asian     & IT                & Daily                                                   \\
P12 & 36  & Female       & College    & Caucasian     & Business Owner                & Weekly                                                   \\
P13 & 33  & Female       & Bachelor & Chinese  & College Staff            & Weekly                                                   \\
P14 & 57  & Female       & College    & African American     & Writer              & Weekly                                                   \\
P15 & 37  & Male       & Bachelor     & Asian     & Developer                & Monthly                                                  \\
P16 & 27  & Male       & College   & African American & Education              & Weekly                                                   \\
P17 & 43  & Trans Male     & College    & White & Unemployed                 & Daily                                                   \\
P18 & 35-40  & Female     & College         & Caucasian & Client Assistant            & Daily                                                   \\
P19 & 43  & Male       & Bachelor    & White & IT                & Weekly                                                   \\
P20 & 18  & Female       & High School    & Caucasian & Student                 & Less than before                                                   \\
P21 & 52 & Male & Bachelor & Chinese\&Irish & IT Manager & Weekly\\
\bottomrule
\end{tabular}
 % \caption{Participant demographic information and other background information.}
 % \xiaozheng{I would suggest following the recommendation for the collection of gender data and reporting it in HCI, e.g., "woman" instead of "female": https://dl.acm.org/doi/10.1145/3338283}
 % \xiaozheng{It would be useful to add another column to Table 1 and clarify the user experience with GenAI. Have the participants used ChatGPT or Bard before (or both)? [please try to mention this in the rebuttal]}
 \label{tab:demographic}
\end{table*}
Our study has some limitations. First, our study explored the mental models of individuals in the United States. The results may not be applicable in other countries or cultural contexts.
Second, our research focused on the chatbot ecosystems, and while GenAI has developed rapidly in recent years, it is still in a phase where the generation of results is unstable. This issue was encountered during our interviews; for example, there were instances where the final response indicating a successful hotel booking could not be displayed, and the travel or personal information requested from different participants varied during the interactions. To mitigate this issue, we prepared a screenshot that shows a successful booking confirmation from Gemini (Figure~\ref{Geminibook}). 
When participants could not complete the booking during the study, we showed them the screenshot instead to ensure a consistent experience. 
It should also be noted that when we conducted the interview, ChatGPT did not support booking hotels directly from its interface even with the Expedia plugin. This is distinctly different from how Gemini works. Yet, it should also be noted that both ChatGPT and Gemini would request users' personal information (i.e., names, room preferences, number of guests, and credit card information) when prompted to book a hotel\footnote{This was the case when the study was complete. At the time of the paper submission, both Gemini and ChatGPT have modified their interfaces to refrain from the request for financial information.}.
\revision{Third, in the ChatGPT portion of this study, the built-in third-party plugin for hotel reservations was Expedia, a well-known booking expert. We have not yet explored how third-party plugins of varying reputations may influence people's mental models, which could be investigated in future work.
Finally, our participants' mental models were influenced by the reputation of the parent companies. In this study, it was our intention to expose participants to all entities involved in the ecosystem to ensure ecological validity. Future work may explore different ways of studying users' mental models without being biased by the company brands.}

\section{Results}

In this section, we present our findings. We first summarized our participants' demographics, then we focused on the participants' four mental models of chatbot ecosystems, their privacy concerns, and their expectations of privacy notice and control in chatbot ecosystems.

\subsection{Participants Demographics}
In total, we have 21 participants. Our participants' ages were between 18 and 62, with an average age of 39. 11 participants were female and 9 were male. Five local participants did the study in-person in our lab while 16 participants did the study via Zoom remotely. 
Our remote participants came from different geographic locations across the US. They also represent a diverse range of occupations, such as university staff, college students, an artistic director, business owners, writers, software engineers, healthcare workers, construction workers, consultants, etc. 
All participants have experience using chatbots. 20 participants have used either Gemini or ChatGPT, and 1 participant has used the AI-powered Microsoft Bing. Participants use chatbots mostly for document editing, ideation (e.g., generating arts or preparing for job interviews), and finding information (e.g., getting recipes or obtaining educational resources). Full demographic information can be found in Table~\ref{tab:demographic}.

\subsection{Mental Models of Chatbot Ecosystems}
\label{mentalmodels}

We identified four mental models, as summarized in Table~\ref{tab:mentalmodeldefinitions}. The four mental models center around the chatbot's role in the chatbot ecosystems and differ primarily on two factors, i.e., the \textit{entities involved in the data flow} and the \textit{perceived trust towards the chatbot}. For better illustration, we named each mental model based on the role of the chatbot.

Specifically, the first three mental models (i.e., Key Player, Medium, Representations) indicate a data flow between the chatbot and its parent company (e.g., ChatGPT’s parent company is OpenAI, and Gemini’s parent company is Google), while only one mental model (i.e., Agent)
involves data flowing directly from the chatbot to the plugins. These results reflect participants' vastly different and diverse understandings of the role of the chatbot in the chatbot ecosystem. Next, we present the mental models in detail.

\subsubsection{Mental Model 1: Chatbot as a Key Player}
\label{subsec:MMOne}

Four participants (P2, P8, P18, P20) held this model when using Gemini. They believed that the chatbot played a key role in the chatbot ecosystem. In the study context, the chatbot \textit{directly and actively assisted} participants' hotel booking by collecting personal information from participants, and then passing their information to its parent company to complete the booking. The Key Player mental model involves three entities: the user, the chatbot, and the parent company. 

\begin{figure}
 \includegraphics[width=\linewidth]{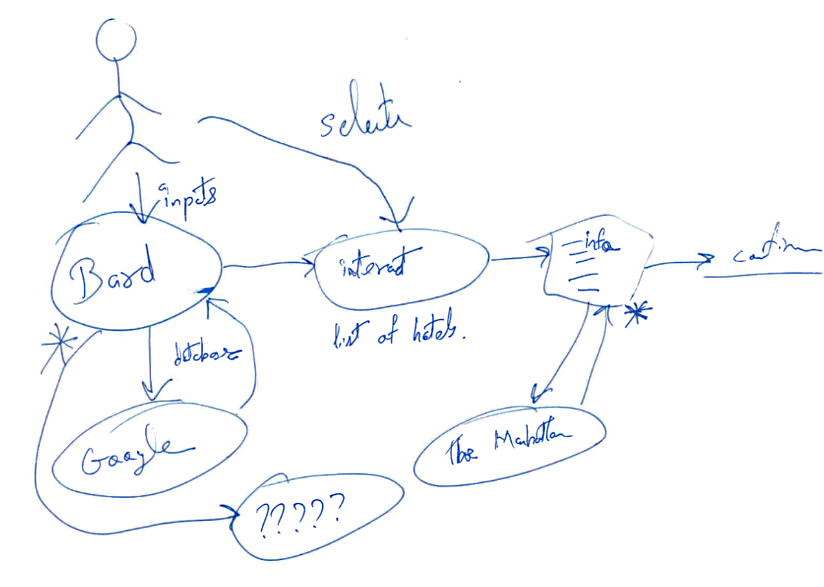}
 \caption{Chatbot as a Key Player Drawing from P2. The data originated from the user and was transmitted to Gemini (Bard), from there it was further transferred to Google, as well as to an entity marked with a question mark, referred to by P2 as the ``black box.'' He believed Gemini collected information and then transmitted it to other uncertain entities other than Google.}
 \label{P2-Gemini}
\end{figure}

For example, P2 (Figure~\ref{P2-Gemini}) believed that he mainly interacted with Gemini throughout the process and provided his information, which Gemini would then pass on to the parent company (Google). He explained, 

\textit{``I guess the first entity that I interacted with is Gemini. And then it took my input, and then, I'm guessing it went to Google... So it went to Google or maybe the database that Google has to get some inputs back. And then I fetch some inputs. And then, it again showed me an interaction. It interacted with me with the list of hotels. And then when I selected one particular hotel, it basically asked me to enter a bunch of information. And then, when I entered that information, it probably went to the hotel booking sites...''} (P2)

\begin{figure}
 \includegraphics[width=\linewidth]{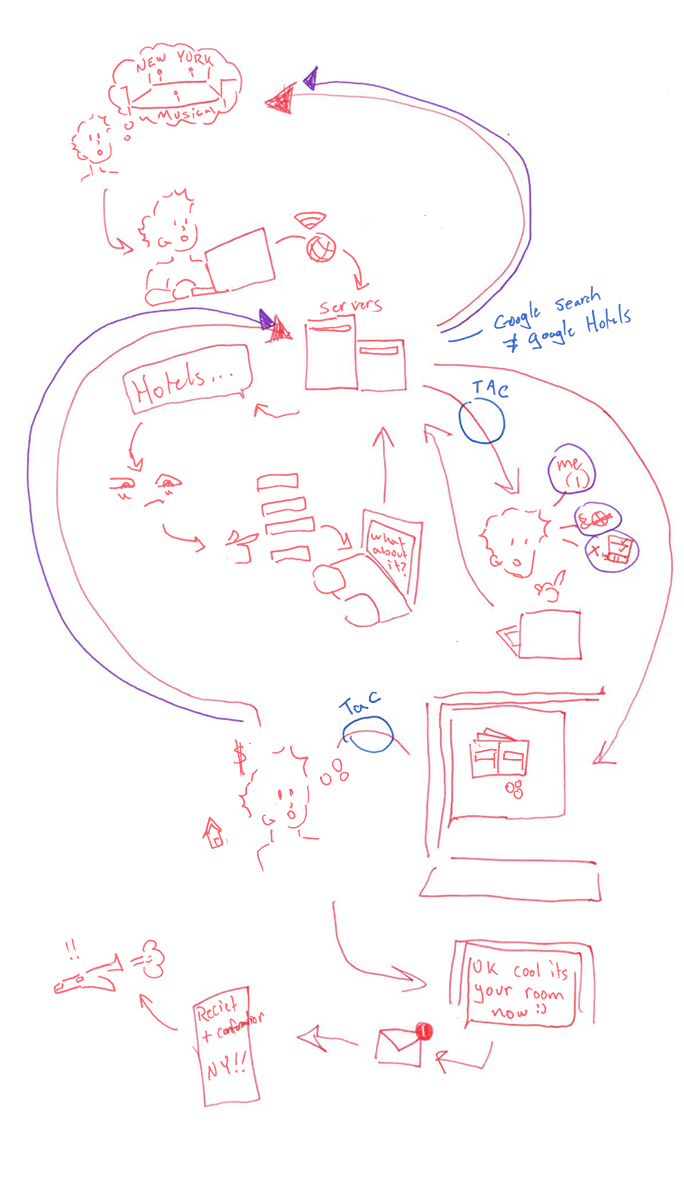}
 \caption{Chatbot as a Key Player Drawing from P20. In P20's drawing, servers were depicted with two rectangles of different sizes, representing the dominant Gemini (Bard) and the subordinate Google. Gemini transferred data to hotels for booking, and it could also be seen that after multiple exchanges within the chatbot ecosystem, the data was transferred back to Gemini's server. }
 \label{P20-Gemini}
\end{figure}

P20 believed that even though the hotel options were retrieved from the parent company's (Google) server to generate hotel options, Gemini still played a key role in the process as it helped process financial information and complete the final booking (Figure~\ref{P20-Gemini}).
% \textit{``...The computer connects to the internet which in turn goes to the servers (Gemini and Google) for generation and the servers start scouring New York for me and tell me about the Hotels. And I look at the list and I choose one, and I ask it to tell me more about that one. And it goes back to the (Gemini) server, and it comes back to me again with its response. And it tells me it needs things like, ``Am I taking people with me?'' ``No, it's just my trip.''...And it goes back through the (Gemini) server again and the (Gemini)server responds. ``Okay, now I need your payment information... I put that in, and I put it in my home address too because that's part of it, I guess. So now knows where I live. It (Gemini)takes that and it buys the room. It then emails it to me, and I have the receipt and the booking confirmation...''}

% \st{We also observed a consistent trend regarding participants' perceived trust towards the chatbot.} 
Among participants who held this mental model, we also observed a consistent trend. That is, participants' perceived trust level was rather independent, meaning that their trust in the parent company did not extend to its chatbot. For example,
P2 believed that Google and Gemini had different levels of rules in handling users' data. He thought that, compared to Gemini, Google had stricter data handling rules and assumed his personal information was stored in Google's possibly more secure database, suggesting that his confidence in Google did not extend to Gemini,

\textit{``I think in my opinion, it's (Gemini) associated with Google, but I think Google has a lot of practices and rules on how they handle their personal information. I'm guessing that they stored my personal information in some database other than bot [Gemini], which might be more secure.''} (P2)

\subsubsection{Mental Model 2: Chatbot as a Medium}
\label{subsec:MMTwo}
\begin{table*}[]
    \centering
     % Reduce font size to fit the table within the page
    \renewcommand{\arraystretch}{1.5}
     \small % Reduce font size to fit the table within the page
    \renewcommand{\arraystretch}{1.5}
    \caption{Summary of the four mental models, their definitions, and factors.}
    \begin{tabular}{p{2cm}p{4cm}p{3.5cm}p{4cm}}
        \toprule
        {\centering \textbf{Mental Model}} & 
        {\centering \textbf{Definition}} & 
        {\centering \textbf{Data Flow Representation}} & {\centering \textbf{Trust towards GenAI Chatbot}} \\ \midrule
        \centering Key Player & The chatbot directly and actively assists in booking hotels. It will transmit information to the parent company and, after obtaining some data from the parent company, return it to the user to finalize the booking. & User - Chatbot - Parent Company & The trust in the parent company does not extend to its chatbot. \\ \midrule
        \centering Medium & The chatbot indirectly assists in hotel bookings by passing information to the parent company. The parent company plays a crucial role in completing the booking. & User - Chatbot - Parent Company - Plugins & Participant's trust in the parent company extends to trusting the chatbot, although participants do not see the chatbot and its parent company as the same entity. \\ \midrule
        \centering Representation & The hotel booking process involves an exchange of information between the user and the parent company, with users perceiving the chatbot as equivalent to the parent company. & User - Parent Company/Chatbot - (Plugins) & Participant's trust in the parent company extends to trusting the chatbot, as they are one entity.\\ \midrule
        \centering Agent & The chatbot transmits the data to the plugin to complete the booking.  & User - Chatbot - Plugins & Participants do not perceive the parent company in the data flow. \\ \bottomrule
    \end{tabular}
    % \caption{Summary of the four mental models, their definitions, and factors.}
    \label{tab:mentalmodeldefinitions}
\end{table*}

Four participants (P1, P3, P12, P17) held this mental model when using Gemini. They believed that in the current chatbot ecosystem, chatbots played the role of a medium and only \textit{indirectly assist} in hotel bookings by passing information to the parent company. As a medium, the chatbot simply offered an interface or a window for participants to connect with the parent company and provide their personal information. Upon receiving the information, the parent company would work with the plugins to complete the booking process. In this process, four entities were involved: the user, the chatbot, the parent companies, and the plugins.

For example, after Gemini passed her information to Google, P3 emphasized information flow between Google and Google Hotels (i.e., first-party plugin), 

\textit{`` I'm going to interface with Gemini, Gemini is connected to Google which obviously collects all of your data anyway. And then... Google is then going to spit out Google Hotels because it wants you to book through there. And I would assume that Google Hotels collects every single bit of information that it can about you... So, it's [Google] gonna then again gather information to spit back out at you and efforts that you'll book through them.} (P3)

\begin{figure}
 \includegraphics[width=\linewidth]{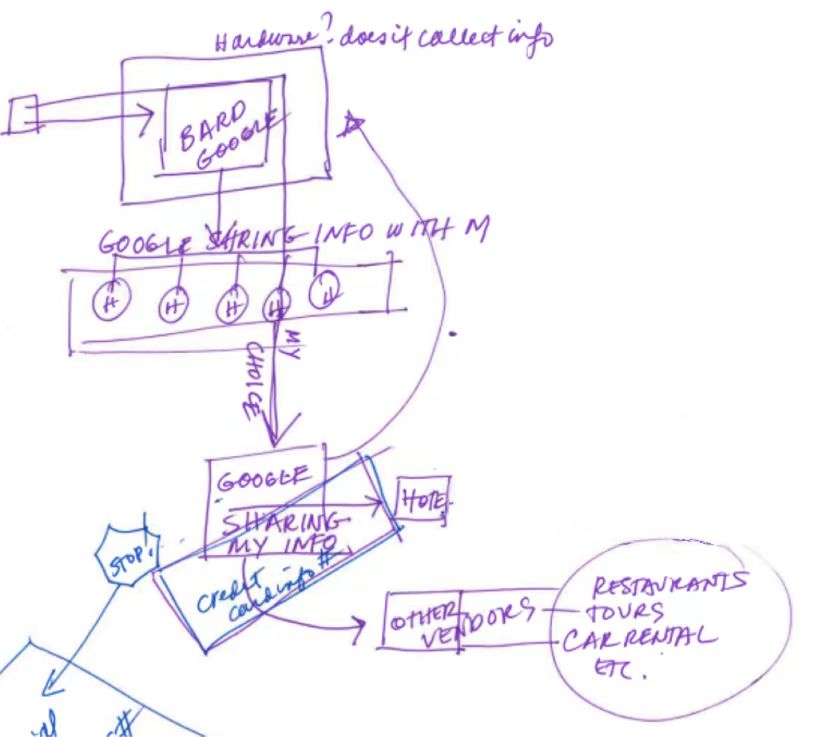}
\caption{GenAI as a Medium Drawing from P1. P1 believed Gemini (Bard) merely acted as a conduit for transmitting information to its parent company, Google. It was Google that actually generated choices for the user and shared information with other vendors.}
\label{P1-Gemini}
\end{figure}

P1 (Figure~\ref{P1-Gemini}) also believed that Gemini could not directly exchange information with hotels for booking; it could only do so by passing the information to Google, and Google would have the capability to book the hotel through Google Hotel,

\textit{``So I am interfacing with them [Gemini], giving them a limited amount of information. They provided me with some information about several different hotels, but the hotels aren't getting any information about me at this point... Then I make a choice, and we have Google sharing my information with the hotel.''} (P1) 

Participants who held this model extended their trust in the parent company to its chatbot, although they did not see the chatbot and its parent company as the same entity. P1 highlighted a tension between the perceived risk of sharing sensitive information with a chatbot and her trust in Google's ability to protect that information, 

\textit{``Right in here where I'm not giving the credit card information directly to the hotel, I'm giving it to the chatbot. And that would be the one [concern] because that's information to share that could have the highest risk. If that piece of information was leaked or somehow the system was compromised, and my credit card information got out there, that could do the most damage to me. But, generally, I trust Google. That's a big company, if I’d share that information, it's safe.''} (P1)

Despite concerns about potential data leaks when sending information to Gemini, P1 felt confident that sharing such information with a Google-associated chatbot was safe due to her overall trust in Google's reputation and security measures. However, when discussing this viewpoint, P1 considered Gemini and Google as separate entities, indicating that she saw them as distinct.

\subsubsection{Mental Model 3: Chatbot as a Representation}
\label{subsec:MMThree}

Participants who held this view believed that the chatbot was a direct representation of its parent company and that the chatbot and the parent company were essentially the same. In other words, the participant's trust in the parent company extended to trusting the chatbot. They saw the chatbot and its parent company as a single entity. When they interacted with the chatbot and shared their personal information, they understood that all the information would be accessible by the parent company.
This mental model involved three entities in its data flow: the user, the parent company (same as the chatbot), and the plugin (if any).

Four participants (P5, P9, P14, P19) held this model when using Gemini. For example, P19 referred to Gemini as ``Google Gemini'' and specifically mentioned that he believed using Google's products was equivalent to using Google itself, and that using any Google product may involve using other Google services as well. He noted Google and Gemini as ``one entity'' as Gemini was developed by Google. Therefore, in his perception, any Google feature, service, or product could be regarded as Google.

\textit{``One entity is the AI, Google Gemini, and then, I would think the other entity is the hotel that you're booking with...So if I'm doing it through Google Gemini and it did everything for me, then I'm only dealing with one entity, which would be Google Gemini...Because it's a product of Google, I'm thinking it’s using Google, so I think it's one...I would assume that any Google feature, any Google search are built into the AI, so I'm putting Gemini AI and Google as the one and the same.''} (P19)

P14 also held the same model. She perceived that Gemini shared data efficiently with Google and believed that the data she shared with Gemini would be used by Google for other purposes in their products. She explained, 

\textit{``Gemini is giving me a few choices of what to look for... Google will sell you anything and everything... I'm gonna just say Google searches for the prompts response... Gemini just responds and it'll say something like, ``I’ve found…'', and it's going to give me a list of hotels... So if you're sharing with Gemini, you're sharing with Google. In the case of someone like Google, obviously, they're storing it (users’ data). They're storing it, and they're using it to sell us more stuff. Quite a few of them [GenAIs] were in beta for a long time, some of them are offering features for free, but no one offers anything for free. And as the saying goes, ``If you're not paying, you're the product.''} (P14)

\begin{figure}
 \includegraphics[width=\linewidth]{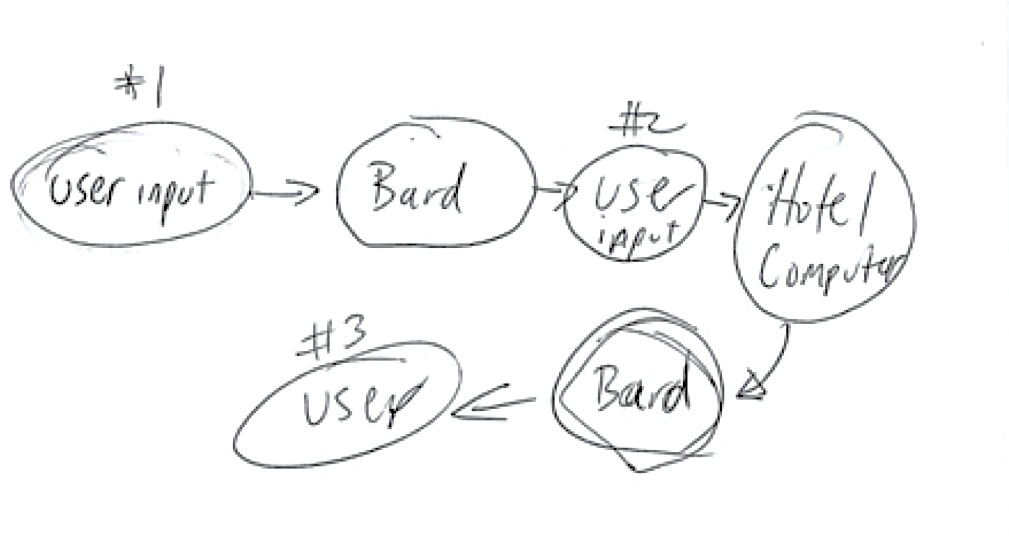}
 \caption{GenAI as a Representation Drawing from P19 (In the diagram, Gemini was used to denote both Gemini (Bard) and Google, with the depiction of Google being omitted. Initially, data was transferred from the user to Gemini. Following an internal data exchange between Gemini and Google, Gemini then acquired more specific information from the user to further the hotel booking process.)}
 \label{Gemini drawing P19}
\end{figure}

Furthermore, in this mental model, since participants perceived chatbots and their parent companies as equivalent, they also directly translated their trust towards the parent companies directly to the chatbots. As such, those who trusted the parent companies would also trust the chatbot, or vice versa. For example, P5 illustrated this by explicitly mentioning his concerns with Google as the reason for not trusting Gemini. 

\textit{``No, I wouldn’t provide [payment information with Gemini)]... because for quite a few years it has been a lot of problems with Google and privacy concerns.''} (P5)

Similarly, P9 expressed that her trust in Google stemmed from her familiarity with it. In her understanding, Gemini was equivalent to Google, so she had the same level of trust in Gemini.

\textit{``You can book directly through Google, so, it probably wouldn't be all that concerned [to share payment information with Gemini], because Gemini is Google. It's a company I'm already familiar with.''} (P9)

\subsubsection{Mental Model 4: Chatbot as an Agent}

\label{subsec:MMFour}

Finally, participants who held the Agent model believed that the chatbot acted as an agent and helped them send queries to first-party plugins (such as Google Hotels in Gemini) and third-party plugins (such as Expedia in ChatGPT) to complete the hotel booking. This process involved three entities: the user, the chatbot, and the plugin (first-party/third-party). The key difference between this model and the other three models was that the chatbot, as an agent, operated independently and played a central role in transmitting information to the plugin (first-party/third-party). As a result, the Agent model did not include the parent company in its data flow and thus, only contained three entities, i.e., the users, the chatbots, and the plugins.

9 participants (P4, P6, P7, P10, P11, P13, P15, P16, P21) held this model when testing Gemini, and, interestingly, all participants held this model when using ChatGPT. We will unpack the comparisons in Section~\ref{comparison}. 

\begin{figure}
 \includegraphics[width=\linewidth]{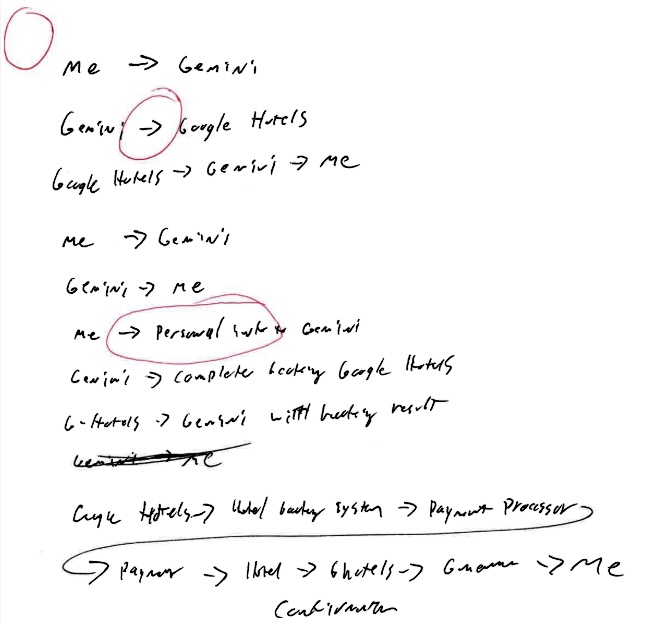}
 \caption{P21's drawing demonstrates how the Agent mental model operated within the Gemini (Bard) Experience. This diagram showed P21's understanding of the bidirectional information exchange between the chatbot and Google Hotels, as well as his view on how Google Hotels processed information and completed the hotel booking.}
 % \xiaozheng{Figure 6 is barely readable. I would regenerate the figure by typing the text beside/below the original text or in the caption.}
 \label{P21-Gemini}
\end{figure}

P21 (Figure~\ref{P21-Gemini}) was an example in the Gemini experience who held the Agent model. He shared his understanding of the information exchange between the chatbot (Gemini) and the Google Hotels plugin,

\textit{``It starts with me entering the query into the chatbot (Gemini), the query getting passed onto the Google Hotels service, and the Google Hotel Service returning the result to the chatbot (Gemini), which it then displays to me. Me asking another query about a specific property to the chatbot (Gemini)... Me, [providing] personal info to the chatbot (Gemini). Then the chatbot (Gemini) uses that to complete a booking through the Google Hotels system, which then returns the result to the chatbot (Gemini), and then it (Gemini)  tells me the details. And then, the chatbot (Gemini) displays that to me. Again, at the high level, at the step-down, obviously Google Hotels is probably authenticating credit cards and stuff like that with payment processors.''} (P21)

P11 believed that Gemini exchanged information with Google Hotels, with Google Hotels collecting the information, returning the booking results to users, and ultimately finalizing the booking. In this process, Google Hotels would have access to his personal information, such as name, contact information, and financial information. He shared,

\textit{``You go to the Gemini website... And then check Google Hotels apparently to see dates, availability, and prices. And then it goes back to the Gemini. And then it asks you which hotel to choose. And then you choose back to Google Hotels after you tell out which dates. And then from there, I would imagine it uses Google Hotelsl for your credit card and all that information. So your name, contact info, like phone and email, your credit card information, and the dates you want. And then I would imagine Google Hotels gets that information.''} (P11)

\begin{figure}
 \includegraphics[width=\linewidth]{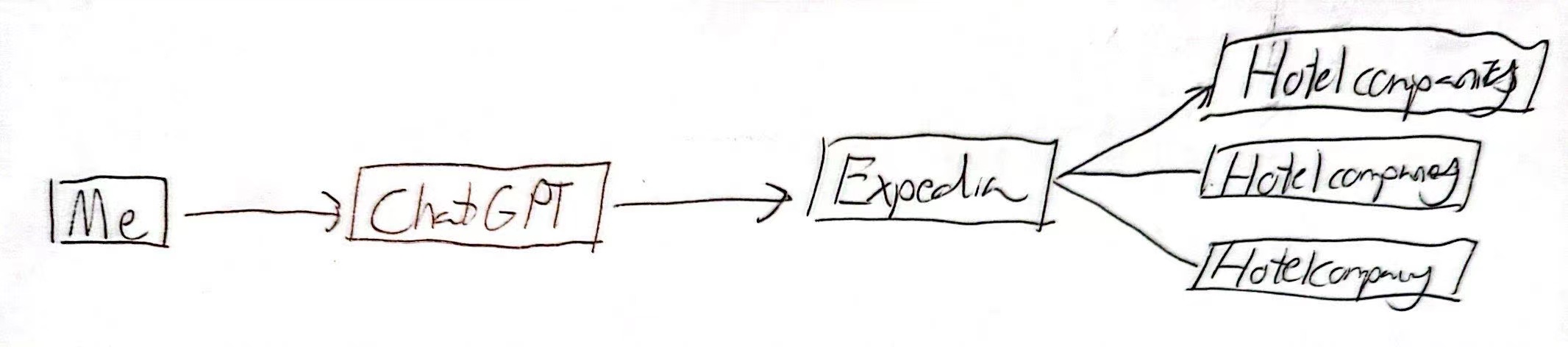}
 \caption{P15's drawing demonstrates how the Agent mental model operated within the ChatGPT experience. The user's data was passed to ChatGPT first, and then Expedia would get the data from ChatGPT to process the booking.}
 \label{P15-gpt}
\end{figure}

Interestingly, all participants who tested ChatGPT held the Agent model as well. For example, P15 (Figure~\ref{P15-gpt}) explained,

% The other two typical examples of Agent MM about ChatGPT experience are P13 and P15.

\textit{``The user and ChatGPT exchange a lot of data and information. And then ChatGPT uses the data from the user to go to Expedia and only Expedia. And then. When it gets information from Expedia, ChatGPT goes back to the user with the information and the link which results in the user going to Expedia as the last step of the chat. So, it's definitely different from what Gemini does. So, the user finishes the last step on it on their own whereas Gemini directly books the rooms. But here, ChatGPT provides the link to the room and then the user is supposed to finish it by themselves...''} (P15)

\begin{table}[t]
\caption{Participants' mental models and whether they have privacy concerns in Bard and ChatGPT. ``Model'' refers to ``Participants' mental models'', and ``Concerns?'' refers to ``Whether participants have privacy concerns''}
\begin{tabular}{c|ccc|ccc}
\toprule
ID & \multicolumn{3}{c}{Bard}              & \multicolumn{3}{c}{ChatGPT}              \\
\midrule
 & Model  & Concerns ? & & Model & Concerns ? \\
\midrule

1  & Medium   & Yes &&  Agent  & No              &                    \\
2  & Key Player &     Yes    && Agent   & No              &                   \\
3  & Medium    & Yes          &                     & Agent  & Yes             &                  \\
4  &  Agent  & Yes          &                   & Agent   & No              &                   \\
5  & Representation   & Yes           &                &  Agent & Yes             &                 \\
6  & Agent   & Yes           &                     &  Agent  & No             &                 \\
7  & Agent   & Yes            &                  &  Agent & Yes              &                   \\
8  & Key Player    & Yes          &                   & Agent  &  No            &                \\
9  & Representation   & Yes           &                    & Agent  & No              &                     \\
10 & Agent   & Yes          &                    & Agent  & No            &                   \\
11 & Agent   & Yes           &                     &Agent  & Yes             &                     \\
12 &  Medium  & Yes           &                    &  Agent  & No              &                     \\
13 & Agent   & Yes           &                    & Agent & No              &                    \\
14 & Representation   & Yes          &                    & Agent  & No              &                     \\
15 &  Agent  & No           &                     & Agent  & Yes              &                     \\
16 & Agent    & Yes          &                    &  Agent & No              &                     \\
17 & Medium   & Yes           &                     &  Agent & No              &                     \\
18 &Key Player    & Yes          &                   &  Agent & No              &                     \\
19 & Representation   & Yes          &                   &Agent   & No              &                     \\
20 & Key Player   & Yes          &                     &Agent    & No              &                   \\
21 & Agent   & Yes          &                     &Agent    & No              &                   \\
\bottomrule
\end{tabular}

\label{tab:participantsmodel}
\end{table}

\subsection{Comparing Mental Models of First-Party and Third-Party Chatbot Ecosystems}

\label{comparison}

We observed some consistency in participants' mental models toward first-party (i.e., Gemini) and third-party (i.e., ChatGPT) chatbot ecosystems. As shown in Table~\ref{tab:participantsmodel}, participants' mental models of Gemini covered all four models, while their mental models of ChatGPT were consistently the Agent model. This suggests that participants' mental models of Gemini are more complicated compared to their mental models of ChatGPT, which essentially consider ChatGPT as an agent that connects users with Expedia via the ChatGPT interfaces. We found this very interesting because the consistency in participants' mental models of first-party and third-party chatbot ecosystems further impacts their perceived concerns. We saw an association between participants' mental models and their privacy concerns towards chatbot ecosystems. Table~\ref{tab:participantsmodel} also provides a high-level summary of participants' mental models and whether they have privacy concerns or not. In general, all but one participant had privacy concerns about Gemini while the majority of participants (16 out of 21) did not have concerns about ChatGPT. This result suggested that our participants had fewer concerns about the third-party chatbot ecosystem compared to the first-party one.

One possible reason for this phenomenon relates to the inherited opaqueness in the chatbots and the broader chatbot ecosystem. Most participants did not understand the working mechanisms of such systems, so when asked about their mental models, they tried to look for different cues to help them understand how the system works. In this process, the third-party system (i.e., ChatGPT) could potentially provide more clues compared to the first-party system (i.e., Gemini). One example is the visual design. Our participants indicated that they had noticed the Expedia icons in the ChatGPT interfaces when prompting ChatGPT to search for a hotel. The Expedia icon clearly indicated that the hotel search was going through Expedia and was not a part of ChatGPT nor its parent company. As a result, participants all believed that their data collected by ChatGPT would eventually be transmitted to Expedia to search for hotels. This perception contributes to the mental model in which the chatbot is considered the agent. P1 explained, 

\textit{``There's me, giving my information to ChatGPT...And ChatGPT is giving my information to Expedia. Expedia provides 3 options and then, I choose, and it still takes me with Expedia, so then it seems like I will leave ChatGPT, at least that's what I saw...And I'm just interacting with Expedia who I give more info to, and then they give it to the hotel.''} (P1)

\begin{figure}
 \includegraphics[width=\linewidth]{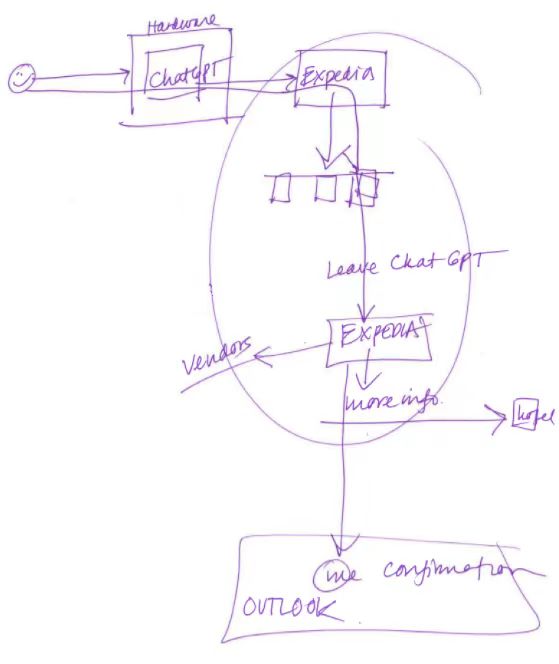}
 \caption{P1's drawing demonstrates how Agent MM Operated Within the ChatGPT Experience (P1 believed that ChatGPT sent the information to Expedia, which then provided multiple hotel options. Additionally, she noted that further information and booking confirmation were all handled by Expedia, specifically mentioning ``leave ChatGPT''.}
 \label{GPT drawing P1}
\end{figure}

In this case, since P1 was very familiar with Expedia for the purpose of hotel booking, she expressed no concerns about it. In comparison, Gemini included the ``Google Hotel'' icon in its interface as well, yet most participants did not notice it in the first place as many considered it part of Google. The highly integrated nature of the Google product ecosystems, to some extent, increases the opaqueness of the chatbot ecosystems and negatively impacts our participants' perceptions. Most participants still considered GenAI something new and did not have sufficient knowledge or familiarity with it. As such, even if they have prior knowledge about their privacy and data usage, they were unable to apply it to the context of the chatbot ecosystem because of such opaqueness. For example, when testing Gemini, P2 compared his experiences with cookie selectors and the chatbots and explained, 

\textit{``I just don't know how my information is handled as supposed to be. Let's say if I enable cookies, I know what's happening, right? I know what's happening there. But here [in Gemini], my information is going in the black box. I don't know if it's going to stay there, or if it's going to come out, or if they're handling it correctly.''} (P2)

The ``black box'' nature of the chatbot ecosystem contributed to P2's uncertainty regarding how his data might be handled. Interestingly, P14 explained a different type of opaqueness when testing Gemini. She noted that, since GenAI has been around for quite some time but without a clear way of making a profit, she believed that her data would be used for advertising purposes. She said,

\textit{``In the case of someone like Google, obviously, they're storing it [users’ data]. They're storing it, and they're using it to sell us more stuff... AI has been around forever in our products, but they opted to just kind of throw everything out there without really having a way to make money as far as I could see... ''} (P14)

In a sense, the highly integrated first-party chatbot ecosystems make it more difficult for users to understand how they work. Additionally, participants' familiarity with third-party plugins and the inherent opaqueness of chatbot ecosystems both contribute to participants' overall concerns towards third-party and more concerns towards first-party chatbot ecosystems. This conclusion is somewhat counter-intuitive and also challenges the persistent higher trust and fewer concerns towards first-party systems in the privacy literature. We will further unpack this point in the discussion. 

\subsection{Privacy Concerns in Chatbot Ecosystems}

When using Gemini, 14 of 15 participants consistently expressed concerns about data processing, sharing, and storage within chatbot ecosystems, aligning with current research on practical privacy issues in GenAI~\cite{zhang2023s, lee2024deepfakes}. Interestingly, most participants did not express the same level of concern regarding ChatGPT. This finding supports our earlier findings in Section~\ref{comparison} and ~\ref{subsec:MMFour}, where participants showed a higher level of trust in ecosystems with third-party plugins and perceived ChatGPT as limited to an \textit{Agent} model. Additionally, compared to existing booking platforms like Booking.com and Expedia.com, participants raised concerns about the maturity and necessity of using chatbots for booking flights and hotels. These doubts contributed to their hesitation to share personal and sensitive information with chatbots. Finally, participants voiced concerns about the lack of adequate laws and regulations specifically governing the privacy practices of chatbots. 

\subsubsection{Privacy Concerns Aligned with Privacy Risks}

Data sharing among multiple entities in the ecosystem raised participants' various types of privacy concerns. Some of our results echo the findings in prior work~\cite{zhang2023s}. We briefly present the repetitive themes as well as the unique ones.

\paragraph{Opaque data practices}

9 participants (P1, P2, P3, P7, P9, P14, P16, P17, P19, P20) mentioned their concerns that the data would used without consent. For example, P1 worried that the information she shared with the chatbot mentioned that the sensitive information she shared with the chatbot could be used for identity theft, particularly involving social security numbers, which could then access financial accounts. Other participants (P1, P3, P9, P14) suspected that the chatbot and its parent company would sell their information to marketing vendors for profit. P2, P7, and P16 voiced concerns about the risk of unauthorized parties accessing their data. They were worried that after sharing information with the chatbot, they would lose control over their data without being aware of who was accessing it. As P9 noted,

\textit{``I would assume just like any other company, if they're [GenAI] not now, they will eventually probably use the [users’] data for advertising or selling the information. I think that's pretty much the norm nowadays.''} (P9)

\paragraph{Surveillance}
6 participants (P3, P9, P14, P17, P19, P20) expressed their concerns about potential surveillance due to the invasive and excessive data collection during interaction with chatbots. For example, P3 talked about her past experiences of having her conversations monitored and her information being collected without her consent. She was concerned that that all her personal information would be collected.

\textit{``[GenAI will gather users' data like] products I use, websites I visit, possibly even what I watch on my TV because I have a smart TV and that is connected to things. So yeah. Things like, my usage of what I watch on TV and maybe even what I eat, maybe even things like my gender identity and sexual orientation. I don't believe any of that is private.''} (P3)

% \subsubsection{Lack of Laws and Policies}

% Lacking laws and public policies to regulate the use of GenAI Chatbot ecosystem is another concern mentioned by our participants. For example, P2 believed that the lack of laws regulating AI might trigger some of her privacy concerns. \textit{``Once we get there [GenAI becomes mainstream], then maybe we'll have better laws and better privacy controls. Because right now there are next to no laws for AI, right?... how it feels is like personal information and stuff, there's like next to no laws.''} (P2)

% \xzrevise{In the case of ChatGPT, the majority of participants did not have privacy concerns as they believed that ChatGPT did not collect personal identifying information during the hotel booking process. However, five participants(P3, P5, P7, P11, P15) had privacy concerns regarding the ChatGPT experience.}

\subsubsection{Concerns About the Maturity and Necessity of Chatbots in Booking Platforms} 

We found that participants believed chatbots, as a new technology, are not as reliable as existing booking platforms (e.g., booking.com, Expedia) when it comes to data management and sharing permissions, particularly for tasks like booking trips—in our case, booking hotels. The belief came from the limited adoption of chatbots for booking purposes, leaving users unfamiliar with their reputation and effectiveness in this area. For example, P18 did not want to share personal and sensitive information on ChatGPT because he had never tried that yet. P2, P4, and P18 noted that chatbots are not commonly used for booking hotels. Their reluctance to share sensitive information, even with reputable company subsidiaries, was due to the technology's novelty.

\textit{``Yeah. It just feels like new technology. Urh, you know, I think if I knew that people in masses were doing it that way, I would be a follower. And I would be willing to provide financial information to Gemini or ChatGPT, but given that still feels new and unknown, I will probably would be a little bit slower to adopt.``} (P4) 

P16 preferred to use established booking agencies because they were more familiar with Expedia. Similarly, P14 questioned the need to share the same information with a chatbot when established agencies already have their details. They preferred sticking with existing services to avoid directly sharing sensitive information with a new system.

\textit{``I've already given my information to Priceline. Why should I go through your ChatGPT when I could go directly to Expedia? My problem (that putting information and booking through chatbot) is just adding more time to a process that used to be pretty simple, you know?``} (P14)

%In our study, only one participant (P15) did not mention any privacy concerns regarding their experience with Gemini, but he mentioned a general concern about the reliability of prompts.

\subsection{Expected Privacy Notice and Control in GenAI Chatbot}

When participants were asked about their expectations for privacy notices and control options in chatbots like Gemini and ChatGPT, most believed that both systems should offer similar privacy-related notice and control. However, there was a distinction in perceptions: while all participants agreed that privacy notices were necessary in Gemini, 6 participants (P1, P6, P9, P13, P19, P20) felt that notices were not essential in ChatGPT, as during the hotel booking process, ChatGPT did not request sensitive information from them.

Our analysis identified several key design considerations regarding privacy notices and controls within the broader chatbot ecosystems, especially concerning data collection, sharing, storage, and usage. Notably, we found no strong correlation between participants’ preferences for privacy notices and their mental models. These insights are elaborated below.

\subsubsection{Privacy Notice}

Most participants viewed privacy notice as essential for informing users about how their data is collected, stored, and used. Participants expressed preferences for a variety of notice formats, including pop-ups, bars, blurbs, emails, disclaimers, and dropdown boxes. About half of the participants also emphasized the importance of user consent in these notices.

\paragraph{Constant Notification} Some participants preferred to receive privacy notices consistently as soon as they entered the chatbot interface. For example, P14 suggested receiving reminders ``every few months'' to inform users that tracking was happening, while P15 wanted a bar displayed at the top of the interface whenever data was being collected. P1 emphasized the importance of a warning displayed within the interface to discourage sharing sensitive information.

\paragraph{Before Data Collection} Most participants preferred receiving privacy notices before the chatbot began collecting personal information, particularly before the system asked for booking details. They favored a concise and clear message explaining how collected data would be used. For instance, participants (P3, P10, P11, P14) expected these notices to be shown after they initiated a hotel booking process or when the chatbot asked for specific booking-related information.

\subsubsection{Privacy Control}

Participants expressed preferences for controlling their data at specific points during their interaction with chatbots. Many indicated that they wanted to configure privacy settings before using the chatbot, while others preferred exercising control during key moments, such as when booking a hotel or after sharing sensitive information.

Some participants also desired the ability to control or delete their data periodically or after certain tasks were completed. For instance, P16 wanted an option for data to be auto-deleted after task completion, allowing users to decide whether data should be retained and for how long, akin to cookie management settings.

Other participants (e.g., P11) wanted a clear system to view and manage their data, allowing them to control which companies or databases had access and to delete data at any time. P14 also emphasized the ability to control data retention duration, drawing parallels to the ease with which cookies can be deleted. Similarly, P15 expressed a desire for control over how their conversations were recorded or stored, suggesting a simple checkbox or toggle to opt out of data collection for review purposes, although they acknowledged that this was not a critical feature for them.

Participants generally preferred intuitive, user-friendly controls, such as pop-ups, checkboxes, and toggles. They emphasized the importance of having control over specific types of data, with options to delete or opt out of data collection. The interface design expectations reflected a preference for minimal disruption during interaction while offering easy-to-use controls. The demand for precise control over specific data types highlighted users’ emphasis on refined management of personal data processing.

\section{Discussion}\label{sec:discussion}
Situating in the large body of literature on users' mental models and the fast development of generative AI, we focused on investigating users' mental models of chatbot ecosystems. We identified four mental models that our participants held based on the role of the chatbot in the chatbot ecosystems. We further found an association between participants' mental models and their privacy concerns, indicating that participants had fewer concerns about third-party systems than first-party systems. 
In this section, we reflect on our results from three perspectives, including the importance of studying users' mental models in chatbot ecosystems, unpacking the comparisons of mental models between first-party and third-party systems, and implications for the design of and policy for future GenAI technologies in general.

% According to Gemini's experience, all entities involved in GenAI EC are considered first-party. In this first-party GenAI EC environment, we have a total of four mental models. However, in the ChatGPT section, we found an interesting shift in the results regarding trust dynamics within the chatbot ecosystem compared to existing literature (can you share some papers for me to cite?). While existing literature suggests that users typically trust first-party entities directly responsible for providing services or technology to them, our research findings indicate that all participants consistently tend to trust third-party entities more. This significant deviation not only challenges the established norms of trust relationships in the digital environment but also paves the way for in-depth exploration of the complex trust formation processes, especially in the GenAI EC domain.

\subsection{Why Do Mental Models Matter in Chatbot Ecosystems?}
\label{discussion:whymatter}

Privacy and security researchers have been studying users' mental models of various technologies and applications, as users' mental models may guide and help reason their behaviors, impact users' attitudes, and inform opportunities for user education~\cite{wash2010folk, yao2017folk, camp2009mental, acquisti2005privacy, asgharpour2007mental, kang2015my}. Our findings in the context of chatbot ecosystems are in line with the prior work. Furthermore, as chatbots and chatbot ecosystems are still relatively new to most general users, their mental models also demonstrate some nuances. For example, users' mental models may have a significant impact on their perceived trust level towards the chatbots and the broader chatbot ecosystems.

As discussed in Section~\ref{mentalmodels}, the perceived relationship between the chatbot and its parent company can significantly impact participants' trust towards the chatbots and the chatbot ecosystem. For example, for participants who held the Representation model, their trust level towards the chatbot was equivalent to that towards the parent company. If they trust the parent company, they would also trust the chatbot. Through investigating participants' mental models of chatbot ecosystems, we were able to clearly identify such relationships (summarized in Table~\ref{tab:mentalmodeldefinitions}). 

As GenAI is evolving from a stand-alone chatbot to an ecosystem (platform) that integrates different internal and external services and features, these mental models and relationships will become increasingly important as they provide a lens for service providers and software developers to come up with ways to enhance users' trust towards their products or correct possible mistrust. Researchers may also leverage the mental models to carry out more precious user education, as risk communications should be tailored towards users' mental models~\cite{camp2009mental}. For example, for those who held the Key Player model, it is useful to focus the education on the data flow between users, chatbot, and the parent companies.

\subsection{Comparing Mental Models in First-party and Third-party Ecosystems}
% \label{first-party VS Thrid-party}
\label{discussion:comparison}

In the privacy and security literature, it is widely acknowledged that users generally have simpler mental models and fewer concerns toward first-party systems compared to third-party ones. However, our results suggested different results in the context of GenAI, as our participants showed four mental models towards the first-party chatbot ecosystem (i.e., Gemini) while only holding one consistent mental model towards the third-party ecosystem (i.e., ChatGPT). This result demonstrates that participants' mental models towards the first-party ecosystem are much more complicated, resulting in more privacy concerns (as summarized in Table~\ref{tab:participantsmodel}). The highly integrated nature of the first-party ecosystem provided fewer opportunities for our participants to understand the data practices of the chatbot ecosystem, and the lack of visual cues that were familiar to our participants further reduced the chances. In the end, the first-party ecosystem created an opaque system that most users did not have prior experiences with, which caused significant concerns. As a comparison, the third-party ecosystem operates in a similar way that users are more familiar with - in a sense, using the Expedia plugin through ChatGPT for hotel booking is somewhat similar to searching for Expedia via Google searches, and most users are familiar with the latter one which they also have less concerns about. 

This comparison, however, does not suggest that our participants had more \textit{accurate} mental models towards the third-party ecosystem compared to the first-party ecosystem. In fact, we believe that all four mental models we identified are either incomplete or inaccurate, as the chatbot ecosystem may include multiple data holders in its various processes~\cite{zhang2023s}. As GenAI is fast developing, users will have to navigate through increasingly complex chatbot ecosystems that involve numerous stakeholders. For example, Microsoft is introducing Copilot in its Office 365, Bing Search, Teams, and other products~\cite{microsoftCopilot2024a, microsoftCopilot2024b, microsoftCopilot2023}; Meta is integrating Meta AI (Llama 3) into its software and hardware product lines~\cite{metaAI2024, Meta2024}. Thus, it becomes critically important to ensure that users have a good and accurate understanding of the working mechanisms of both first-party and third-party chatbot ecosystems so that they can make informed decisions about their privacy and data.

Next, we will discuss the design and policy implications based on our findings.

\subsection{Implications for Design and Policies}

As GenAI is increasingly integrated into our societal infrastructure, it calls for joint efforts from developers, policymakers, and scholars to reevaluate the design and regulation of GenAI services, stressing the need for transparency, educative initiatives for users, and the implementation of stringent privacy measures across all involved entities.

\textbf{Design implication: Provide transparency features in chatbot ecosystems.}
As mentioned in previous sections, one critical challenge is the inherent opaqueness of chatbot ecosystems. General users typically do not have knowledge of how the system works and how their data will be transmitted and used. We suggest that chatbot ecosystems should incorporate transparency features to help users understand the mechanisms of ecosystems. One existing example is the Expedia icon in ChatGPT when prompting it to search for a hotel. Future designers should explore other transparency mechanisms, such as a visualization in the chatbot interface to show how their data flows among different entities, a tool that automatically shows the impact on users' privacy when sensitive data input is detected, etc.
% \xiaozheng{C: The paper could do more to explore the specifics of GenAI that raise unique privacy concerns, such as the potential use of input data for training, and the limited explainability of responses. While mentioned in 2.1, that section didn't get follow-through in discussion of the interview results. Especially when a technology is new to its public, users will make analogies to more familiar tools. How well do these analogies match? What alternative explanations could help to make the roles and functions clearer? }

\textbf{Policy implication: Regulate the disclosure of involved entities in chatbot ecosystems.}
At the time of this research, we studied the privacy policies of various GenAI platforms and found that most privacy policies focus on the types of data being collected and how the data is used. However, our results suggest that the \textit{involved entities} may have a significant impact on users' perceived trust and privacy concerns towards GenAI. As such, we recommend that public policies and regulations should require the disclosure of involved entities in chatbot ecosystems. Such a requirement would echo the call for transparency in the previous section from the policymakers' perspective and will enhance users' understanding of the working mechanisms of GenAI, which will eventually help them make informed decisions about their privacy. 

\textbf{Research direction: Understand how to leverage third-party entities to gain user trust.}
Our participants demonstrated a higher level of trust and fewer privacy concerns towards the third-party chatbot ecosystem due to its visual design, users' perceived involved entities, and familiarity. It is possible, however, that other underlying causes also exist behind users' preference for third-party services. Future research may investigate whether users' preferences over third-party GenAI services are consistent throughout different third-party platforms; and if so, what specific characteristics or actions by these third-party services cultivate a higher degree of trust among users. The resulting insights can offer valuable guidance for creating GenAI systems that are inherently more trustworthy, alongside informing educational initiatives that bolster user confidence and security.

\section{Conclusion}
Generative Artificial Intelligence (GenAI) has rapidly developed in recent years. As GenAI's capabilities start to greatly expand, it is turning into a platform infrastructure, i.e., GenAI Ecosystems. In this research, following the abundant literature in privacy and security research, we adopted a mental model approach and investigated users' understanding of how GenAI ecosystems work. Through 21 semi-structured interviews, we uncovered users' four mental models towards first-party (e.g., Google Gemini) and third-party (e.g., ChatGPT) chatbot ecosystems. These mental models centered around the role of the chatbot in the entire ecosystem. We further found that participants held a more consistent and simpler mental model towards third-party ecosystems compared to the first-party ones, resulting in their higher level of trust and fewer concerns towards the third-party ecosystems. We discuss the design and policy implications based on our results. 

\section{Acknowledgment}
We thank the reviewers for their invaluable feedback and our participants for supporting this research. This research is in part supported by NSF CNS-2426397, NSF CNS-2232653, and a Google PSS Faculty Award. We also thank Allen Yilun Lin, Chaoran Chen, and Toby Jia-Jun Li for their thoughtful feedback. 

\bibliographystyle{ACM-Reference-Format}
\bibliography{main}

% \appendix
\section{Appendix}

\section*{Interview Protocol}

\subsection*{General Questions}
Here, we want to give you our definition of ``GenAI Chatbots.'' They are large language model-based, trained on a massive dataset of texts and code, and can perform many kinds of tasks given by users. There are a couple of chatbots in the market now such as ChatGPT, which has been developed by OpenAI, and Bard by Google. Do you have any questions about this? Okay, now, let's continue.

\begin{enumerate}
    \item You’ve told us in the survey that you have experience with GenAI chatbots. What would you say your familiarity with them is?
    \begin{enumerate}
        \item What chatbot have you used before?
        \item When and why did you use them?
        \item How often do you use them?
        \item What do you think of them?
    \end{enumerate}
    \item Have you ever used GenAI chatbots to search for services, such as booking hotels, flights, etc.?
    \item Could you share a recent instance where you asked ChatGPT or a similar chatbot for advice on a purchase or reservation?
    \item GenAI chatbots such as ChatGPT have plugins that provide additional services to users, such as Expedia.
    \begin{enumerate}
        \item How familiar are you with these plugin features?
        \item What is your experience with these features?
    \end{enumerate}
    \item When you use the GenAI chatbots, were there any cases in which you have to share some information with them?
    \begin{enumerate}
        \item For example, your personal information, your preferences, habits, plans, and other things?
        \item (If yes) What did you share? Anything you did not share? Why or why not?
        \item (If no) Why not?
    \end{enumerate}
    \item How do you think GenAI chatbots use your data? If so, which kinds of data do you believe they might be using?
\end{enumerate}

\subsection*{Drawing Exercises}

Before the drawing exercise starts, you'll have two hands-on experiences with prominent GenAI chatbots—Bard and ChatGPT, which have been developed by Google and OpenAI as we introduced at the beginning.

Now, we will give you a scenario. Imagine you are going on a trip, and you want the GenAI chatbot to help you book a hotel near 10019 for Dec. 21st to 22nd in New York City. Try to interact with the GenAI chatbot by using these prompts we printed out for you. Throughout the entire process, we will use our own account, so we won't make any use of your personal information. Do you have any questions? OK, just go ahead.

\textbf{(Provide Information)}

(Once they are done on Bard) OK, now, you have stopped here. Since Bard is still in the testing phase, the prompts vary each time, but most of the time, it successfully helps you book a hotel. I can show you screenshots of successful bookings. [insert the screenshot]

Next, we want you to do a drawing task. Think about the process you just went through. The process is—you want to find a hotel and ask the chatbot to book it for you, you provide your information, and the hotel is booked. I want you to draw this process on this piece of paper. More importantly, I want you to focus on a couple of things: 
\begin{enumerate}
    \item Which entities are involved in this process?
    \item How your data is collected, transmitted, and used by these entities from the moment you log in, to the moment you complete the booking. While drawing, we strongly encourage you to think aloud, feel free to share your thoughts with us.
\end{enumerate}

Please go ahead. Please let me know if you have any questions.

After the participant has finished the diagrams and explained their ideas, we will continue to ask the following questions.
\begin{enumerate}
    \item Can you explain your diagrams to me?
    \item Are there any components in your drawing that concern you?
    \begin{enumerate}
        \item For example, any entity, any particular data collection and data flow, etc.
    \end{enumerate}
    \item How comfortable do you feel sharing personal information with GenAI chatbot to get more satisfactory generated results?
    \begin{enumerate}
        \item Which parts of the diagram are you comfortable with?
        \item Which parts are you not comfortable with?
    \end{enumerate}
    \item In your drawing, thinking about the data collection, would you say notifying you about the data collection is necessary?
    \begin{enumerate}
        \item [If yes] How would you want to be informed about data collection when using GenAI chatbot for this kind of service? At which step do you prefer to be informed? You can label it in the diagram if necessary.
        \item [If no] Why not?
    \end{enumerate}
    \item You mentioned that you would like some kind of notice/consent/control here. How would it look like? I’d like to ask you to draw on this piece of paper an ideal notice interface for you.
    \item (If their drawing or answers do not contain any type of control) In your drawing, you mentioned some kind of notice and you showed us how it may look like. I’m curious to hear whether you would like to have any kind of control. Let’s say you have the superpower to control your data flow in this ecosystem, what kind of control do you like to have? Where do you like it to happen? Can you point it out in the diagram?
\end{enumerate}

After the first part is finished, we will let the participant try the second one and do the same process.

Next, we'd like you to use ChatGPT to complete the same process—book a hotel near 10019 in New York for December 10th to 12th. You can now begin interacting with ChatGPT.

% After users click on the external link, make sure to remind them to be cautious as their entered dates might be directly transferred to a third-party entity by ChatGPT.

% \begin{enumerate}
%     \item Can you explain your diagrams to me?
%     \item Are there any components in your drawing that concern you?
%     \begin{enumerate}
%         \item For example, any entity, any particular data collection and data flow, etc.
%     \end{enumerate}
%     \item How comfortable do you feel sharing personal information with AI chatbot to get more satisfactory generated results?
%     \begin{enumerate}
%         \item Which parts of the diagram are you comfortable with?
%         \item Which parts are you not comfortable with?
%     \end{enumerate}
%     \item In your drawing, thinking about the data collection, would you say notifying you about the data collection is necessary?
%     \begin{enumerate}
%         \item [If yes] How would you want to be informed about data collection when using AI chatbot for this kind of service? At which step do you prefer to be informed? You can label it in the diagram if necessary.
%         \item [If no] Why not?
%     \end{enumerate}
% \end{enumerate}

\subsection*{Demographic Information}
\begin{enumerate}
    \item What do you do for a living?
    \item How old are you? If you'd rather not give a specific number, could you kindly indicate an age range or group you belong to?
    \item How do you identify your gender?
    \item Which ethnic or cultural group do you most closely associate with?
    \item What’s your highest level of education?
\end{enumerate}

% \xiaozheng{A: One of the participants mentioned that they noticed the Expedia icons in the ChatGPT interface (line 1084), and later, the authors discussed that such visual clues might have contributed to transparency (line 1287). It would be useful to share the screenshots of Bard and ChatGPT (in the way they appeared for the participants) as an appendix and highlight the differences between the two designs.}

%\input{sections/Rebuttal}

\end{document}